\documentclass[prc,superscriptaddress,unsortedaddress,twocolumn]{revtex4}

\usepackage[english]{babel}
\usepackage{graphicx,epsf,bm,amsmath,color}
\usepackage{here,ulem}
\def\bk{\mbox{\boldmath $k$}}
\def\bp{\mbox{\boldmath $p$}}
\def\bq{\mbox{\boldmath $q$}}

\def\cg0m#1#2#3{\( ( #1 0 #2 0 | #3 0 ) \)}
\def\cg0#1#2#3{( #1 0 #2 0 | #3 0 )}

\def\h3t{\mbox{$_\Lambda^3$H}}

\begin{document}
\title{Faddeev calculations of low-energy $\Lambda$-deuteron scattering\\
and momentum correlation function}
\author{M. Kohno}
\affiliation{Research Center for Nuclear Physics, Osaka University, Ibaraki 567-0047,
Japan}


\author{H. Kamada}
\affiliation{Department of Physics, Faculty of Engineering, Kyushu Institute of Technology,
Kitakyushu 804-8550, Japan}
\affiliation{Research Center for Nuclear Physics, Osaka University, Ibaraki 567-0047,
Japan}
\begin{abstract}
Faddeev calculations of low-energy $\Lambda$-deuteron elastic scattering are performed
up to $E_{cm}=20$ MeV crossing the deuteron threshold. The phase
shifts of the $s$ wave with $J=1/2$ and $J=3/2$ are calculated using strangeness $S=-1$
hyperon-nucleon interactions in chiral effective field theory NLO13 and NLO19 parametrized by the
J{\"u}lich-Bonn group. The effective range parameters, specifically the
scattering length and effective range, are ascertained
through the calculated phase shifts. The present study evaluates the momentum
correlation functions of the $\Lambda$-deuteron system using the $\Lambda$-deuteron relative
wave function, constructed from half-off-shell $t$ matrices. The results are then compared with
those obtained using an approximate formula.
\end{abstract}

\maketitle
\section{Introduction}
An accurate description of $\Lambda$-nucleon interactions is essential for achieving a
microscopic understanding of $\Lambda$ hypernuclei.
This understanding is being sought through the collection of increasingly accurate
experimental data at several facilities \cite{TAM22}.
Furthermore, this knowledge is also crucial
to understanding the role of $\Lambda$ hyperons in neutron-star matter.
Although various theoretical descriptions of the $\Lambda$-nucleon interaction
have been developed over the past several decades,
the quality of these models remains comparatively
inferior to that of $NN$ potentials, largely due to the scarceness of available scattering data. 
The absence of the $\Lambda N$ two-body bound state is a significant
disadvantage. Then, the lightest hypernucleus, \h3t, plays a crucial role in investigating
the $\Lambda$N interactions to determine the $s$-wave strength, despite
the difficulty in controlling the relative ratio of singlet and triplet channels.
However, the shallow separation energy of \h3t to $\Lambda$ and deuteron has not been
adequately determined and has a considerable error bar.
The current world average is $164\pm 43$ keV  \cite{MA2023}.
There is a problem with the presence
of $\Lambda NN$ three-body forces (3BFs) on the theoretical side.
It is crucial to understand the role of the 3BFs quantitatively. We have carried
out Faddeev calculations for \h3t \cite{KKM22} using next-to-leading order (NLO)
hyperon-nucleon ($YN$) interactions and  $YNN$ 3BFs provided
by the expressions in the next-to-next-to leading order (NNLO) in chiral effective
field theory (ChEFT). The net contribution of the 3BFs is not negligible but is
of a similar magnitude as the present experimental
uncertainty. However, the result depends
on the low-energy constants which are difficult to fix without investigations
of heavier hypernuclei. After observing the order
of magnitude of the 3BF effect in the bound \h3t, it is worthwhile to study the role
of $\Lambda$N interactions in scattering processes.

There are several theoretical studies in the literature on the properties
of the $\Lambda d$ scattering. Garcilazo \textit{et al.} \cite{GAR75, GAR76} used
bound state Faddeev calculations to estimate effective range parameters.
Hammer  \textit{et al.} \cite{HWH02,HH19} carried out studies within the
framework of pionless effective field theory. Sch\"{a}fer  \textit{et al.} \cite{SCH22}
discussed the $J=3/2$ $\Lambda d$ phase shift in low energies based on varying
effective range parameters. However, no explicit calculation of the $\Lambda d$
elastic scattering has been performed by solving Faddeev equations
using modern $YN$ interactions.

In this article, we consider $\Lambda$-deuteron scattering problems
in a Faddeev formulation using chiral $NN$ and $YN$ interactions, although
3BFs are not incorporated. While the data from direct low-energy $\Lambda d$
scattering experiments are still forthcoming, the $\Lambda$d correlation functions
observed in heavy-ion collision experiments \cite{FAB21} have already provided valuable
insights into the interactions between $\Lambda$ and
the deuteron as well as between $\Lambda$ and the proton. The preliminary results of the
$\Lambda d$ correlation functions were reported in a report \cite{HU24}
based on 3 GeV Au+Au collisions by the STAR collaboration.
The correlation function depends on the $\Lambda d$ relative wave function,
which is controlled by the interaction between $\Lambda$ and deuteron arising from
$\Lambda N$ interactions controls. 
Thus far, theoretical investigations of the $\Lambda d$ correlation function have
been carried out \cite{HAID20} using an asymptotic wave function approximated
by effective range parameters of the scattering amplitude \cite{LL82}.
It is worthwhile to calculate the corresponding correlation function using
the $\Lambda d$ wave function obtained by solving the Faddeev equation
for the $\Lambda$ scattering process on the deuteron.

The deuteron is a two-body composite system, and thus the momentum
correlation function is affected by the dynamics associated with its formation.
The problem was studied by Mr\'{o}wozy\'{n}ski \cite{MR20} and actual calculations
for the nucleon-deuteron correlation functions were performed by Viviani \textit{et al.}
\cite{VIV23}. Similar calculations for the $\Lambda d$ case are possible in the
present Faddeev treatment of the $\Lambda d$ scattering, which is a subject of future
investigation.

The Faddeev equations for the $\Lambda$-deuteron ($\Lambda d$) scattering
are outlined in Sec. II, based on the formulation
by Gl{\"o}ckle \textit{et al.} \cite{GL96}. The equations in a
partial wave representation and the treatment
of two types of singularities, a moving singularity and a deuteron pole,
are outlined in Appendices A$-$C.
The numerical results for the $s$-wave phase shift up to $E_{cm}=20$ MeV are
presented in Sec. III. The parameters of the effective range expansion, i.e.,
the scattering length and effective range, are estimated
based on the obtained phase shifts. The radial $\Lambda d$ wave function is
evaluated and $\Lambda d$ correlation functions corresponding to the
measured ones in heavy-ion experiments are calculated.
A summary is provided in Sec. IV.

\section{Faddeev equations for $\Lambda$-deuteron scattering}
We follow the derivation of the equations describing three-body scattering problems in Ref. \cite{GL96}.
Two nucleons are denoted by labels 1 and 2. The $\Lambda$ hyperon is assigned
label 3. The mass difference of the proton and neutron is discarded. In the present treatment,
 three-body $\Sigma NN$ states are not incorporated explicitly, although the $\Lambda N$-$\Sigma N$
coupling is included in solving $\Lambda N$ $t$ matrices. Three-body interactions are not considered.

The process of the $\Lambda$-hyperon scattering on
the bound deuteron is described by the following set of equations:
\begin{align}
 u_3\phi=& t_1 G_0 u_1 \phi + t_2 G_0 u_2 \phi, \\
 u_1\phi=& G_0^{-1} \phi + t_3 G_0 u_3 \phi + t_2 G_0 u_2 \phi, \\
 u_2\phi=& G_0^{-1} \phi + t_3 G_0 u_3 \phi + t_1 G_0 u_1 \phi,
\end{align}
which correspond to Eqs. (36)$-$(41) of Ref. \cite{GL96} with the replacement
as $\phi_1 \rightarrow \phi$,
$U_{11}\rightarrow u_3$, $U_{21}\rightarrow u_1$, and $U_{31}\rightarrow u_2$.
$G_0$ is a three-body Green function and $t_i$ is a pertinent two-body $t$ matrix.
Because the two-nucleon state is antisymmetric, Eqs. (2) and (3) are
equivalent, and the above equations reduce to 
\begin{align}
 u_3\phi=& (1-P_{12}) t_2 G_0 u_2 \phi, \\
 u_2\phi=& G_0^{-1} \phi + t_3 G_0 u_3 \phi -P_{12} t_2 G_0 u_2 \phi,
\end{align}
where $P_{12}$ denotes the exchange operation for nucleons 1 and 2.
In actual calculations, it is convenient to introduce $T_i \equiv t_i G_0 u_i$.
The above equations are rewritten as follows:
\begin{align}
 T_3\phi=& t_3 G_0 (1-P_{12})T_2 \phi, \label{eq:pe1}\\
 T_2\phi=& t_2 \phi + t_2 G_0 T_3 \phi - t_2 P_{12} G_0  T_2 \phi. 
\label{eq:pe2}
\end{align}

\begin{figure}[b]
\centering
\includegraphics[width=0.3\textwidth]{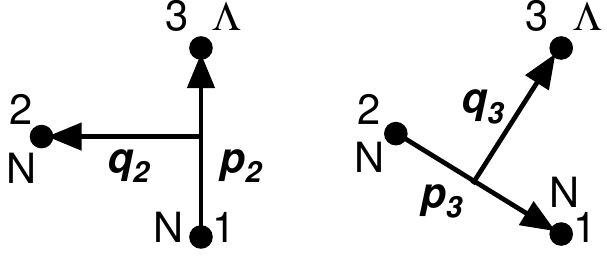}
\caption{Jacobi coordinates.
}
\label{fig:jacobi}
\end{figure}
These equations are solved in a partial wave representation in momentum space.
Two sets of the Jacobi momenta $(\bp_j,\bq_j)$ with $j=2,\;3$ are defined
as in Fig. \ref{fig:jacobi}. The partial wave three-body state in a Jacobi momentum
space is denoted as
\begin{align}
 |p_j q_j\alpha_j \rangle =& |p_j q_j ; [[\ell_{p_j} \times s_{p_j}]^j\times
 [\lambda_{q_j}\times s_{q_j}]^I]^J_{M}, \notag \\
 & [t_{p_j}\times t_{q_j}]^T_{M_T}\rangle,
\label{eq:pw}
\end{align}
where an abbreviated notation for an angular-momentum coupling with Clebsch-Gordan
coefficients is used:
\begin{equation}
 [\ell \times s]^j_{m_j}=\sum_{m_\ell m_s} (\ell m_\ell s m_s|j_{m_j})
 Y_{\ell m_\ell}(\hat{\bp}) \chi_{s m_s}.
\end{equation}
Here, $Y_{\ell m_\ell}$ is a spherical harmonic function, and $\chi_{s m_s}$ represents a spin state.  

By solving the simultaneous equations (6) and (7), the $T$ matrix of the elastic
$\Lambda$-deuteron scattering is obtained \cite{GL96} by
\begin{equation}
 T=\langle \phi|(1-P_{12})T_2|\phi\rangle =2 \langle \phi|T_2|\phi\rangle 
\end{equation}
The $S$-matrix is related to the $T$-matrix as
\begin{equation}
 S=1-2\pi i \mu_{\Lambda d} kT, 
\end{equation}
where $\mu_{\Lambda d}$ is a reduced mass of the $\Lambda$ and deuteron, and $k$
is the $\Lambda$-deuteron relative momentum. 

Explicit equations in a partial wave representation are presented in Appendix A.
In solving these equations numerically, there are difficulties in treating two types
of singularities. One appears in the Green function $G_0$, which is known
as a moving singularity. This singularity is treated by a standard subtraction
method in evaluating the matrix elements. The other one is the
deuteron pole, which has to be taken care of when $T_3$ is substituted in
Eq. (7). The way to treat the deuteron pole and the moving singularity is
outlined in Appendicies A and B.

\section{Calculated results}
\begin{figure}[b]
\centering
\includegraphics[width=0.45\textwidth]{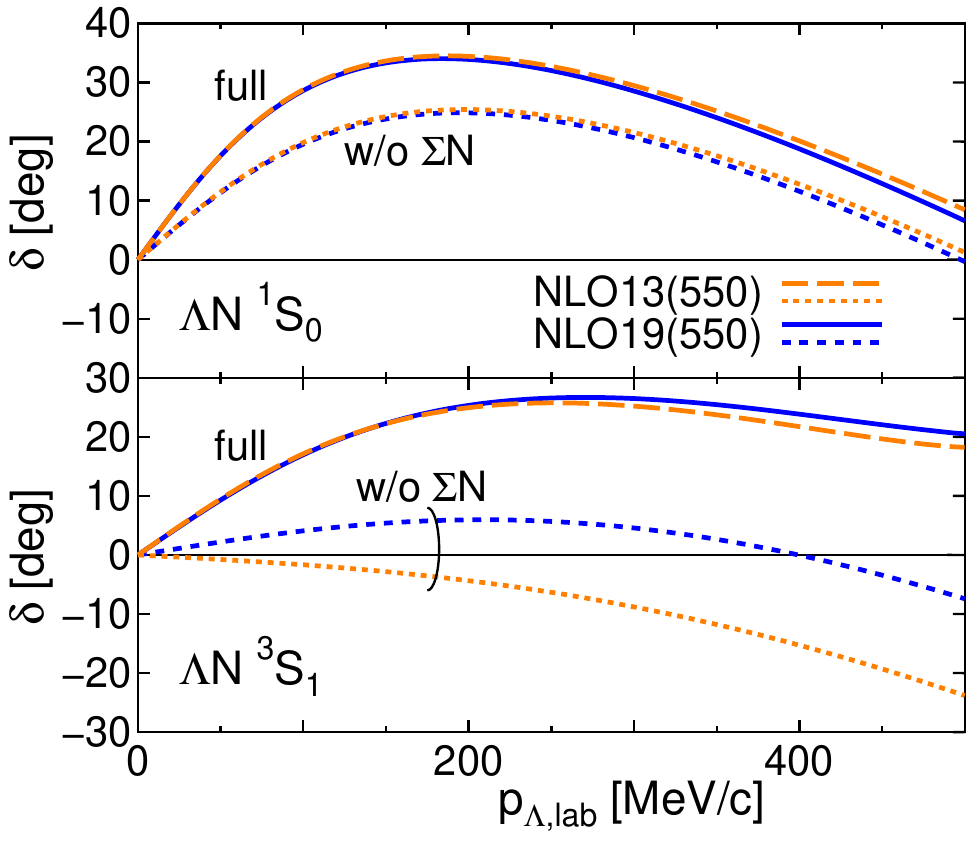}
\caption{Phase shifts of $\Lambda$-nucleon elastic scattering in  $^1$S$_0$ and $^3$S$_1$
channels with chiral NLO13 and NLO19 interactions as a function of the incident laboratory
momentum. Two interactions yield almost identical results.
Results with switching off the $\Lambda$N-$\Sigma$N coupling are included. 
}
\label{fig:ln}
\end{figure}
In this section, we present the results of the Faddeev calculations for low-energy
$\Lambda$-deuteron elastic scattering in an $s$ wave up to $E_{cm}=20$ MeV
using two versions of chiral NLO $S=-1$ $YN$ interactions, NLO13 \cite{NLO13} and NLO19
\cite{NLO19}. As for the $NN$ interaction, the chiral N$^4$LO$^+$ interaction \cite{RKE18} is used.
The cutoff scale is 550 MeV for both $NN$ and $YN$ interactions.

Before presenting the results of the $\Lambda d$ scattering, it is instructive to show
$^1$S$_0$ and $^3$S$_1$ phase shifts of the $\Lambda N$ scattering with NLO13
and NLO19. Figure \ref{fig:ln} shows those results.  Phase shifts obtained
by switching off the $\Lambda N$-$\Sigma N$ coupling are included to show the
difference between the characters of two interactions. 

\subsection{$s$-wave phase shifts of $\Lambda d$ scattering}
In the Faddeev calculations of the $s$-wave phase shifts of  the low-energy
$\Lambda$-deuteron scattering presented below, the orbital angular momenta
$\ell_{p_j}$ and $\lambda_{q_j}$ are restricted to zero for both $j=2$ and $3$ to
reduce the computational load,
although the tensor coupling is naturally included in evaluating $NN$ and $\Lambda N$
$t$ matrices. 
We have checked that the inclusion of the states with $\ell_{p_j}=2$ and
$\lambda_{q_j}=2$ has a very small effect on the $s$-wave phase shifts.

\begin{figure}[b]
\centering
\includegraphics[width=0.48\textwidth]{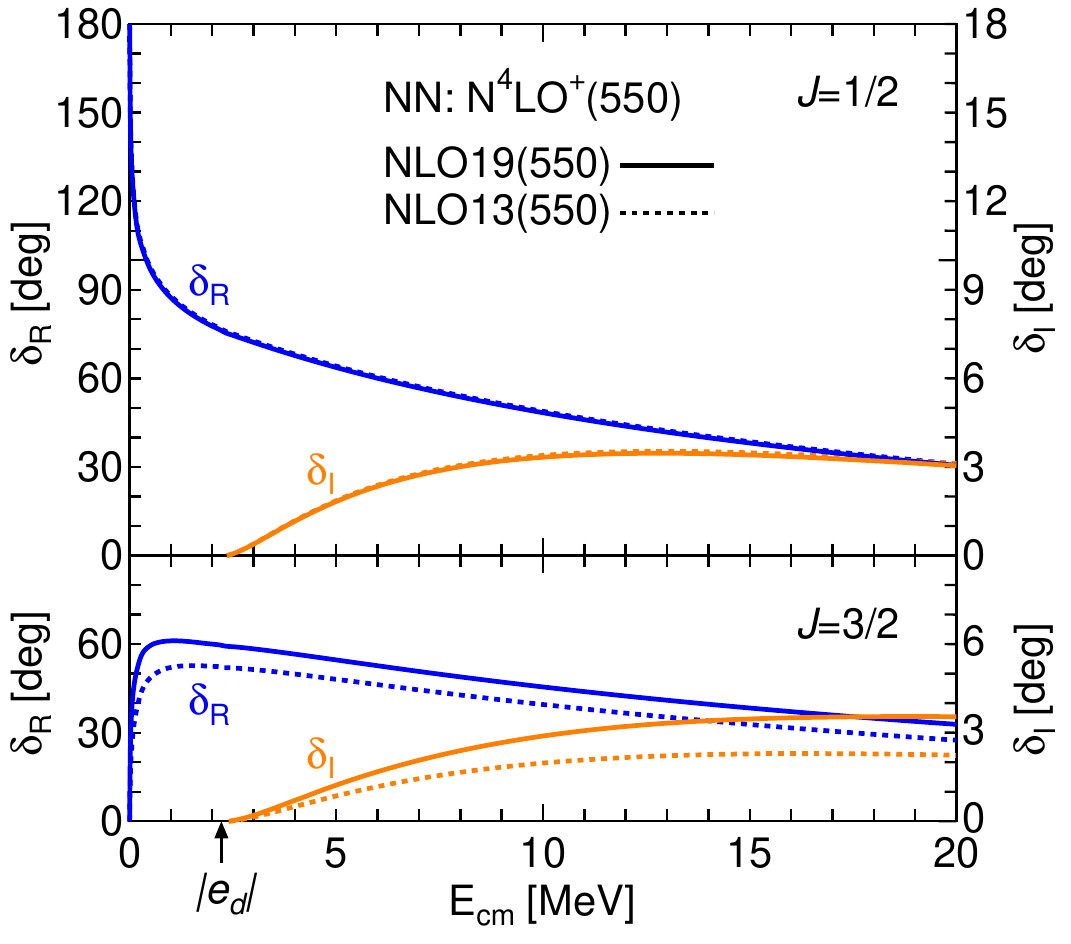}
\caption{Calculated $\Lambda$-deuteron $s$-wave phase shifts as a function of $E_{cm}$.
The left (right) vertical scale is for the real (imaginary) part. 
The upper (lower) panel shows the phase shifts in the $J=1/2$ ($J=3/2$) channel. The solid (dashed)
curves are for the chiral NLO19 (NLO13) $YN$ interaction of the J{\"u}lich-Bonn
group with the chiral N$^4$LO$^+$ $NN$ interaction. The solid and dashed curves are almost
indistinguishable in the case of $J=1/2$. The cutoff scale is 550 MeV for both $NN$
and $YN$ interactions. }
\label{fig:phase}
\end{figure}

The upper panel of Fig. \ref{fig:phase} shows the real and imaginary phase shifts
of $\Lambda d$ elastic scattering in the $J=1/2$ channel
for both NLO13 and NLO19 $YN$ interactions. The imaginary part appears at the
deuteron breakup threshold $E_{cm}=|\epsilon_d|$, where $E_{cm}$ is the incident
center-of-mass energy and $|\epsilon_d|$ is the deuteron binding energy.
Due to the presence of the bound hypertriton in this channel,
the real part of the phase shift starts from 180$^\circ$. Because the NLO13 and
NLO19 interactions are tuned to describe the binding of \h3t, the calculated results
are almost indistinguishable.

The lower panel of Fig. \ref{fig:phase} represents the real and imaginary phase shifts in the
$J=3/2$ channel for both NLO13 and NLO19 $YN$ interactions. Because the $^1$S$_0$
$\Lambda$N interaction is irrelevant for $J=3/2$, the difference of the phase shifts
between NLO13 and NLO represents different properties of the $^3$S$_1$ part of these
interactions, despite the almost identical phase shifts of the $^3$S$_1$ phase shifts shown
in Fig. \ref{fig:ln}. The behavior of the phase shifts corresponds to the absence of a bound
state in this channel. The enhancement of the cross-section at low energies, however,
implies the presence of a pole close to the axis. The position of the virtual state is
approximated by the effective range parameters as \cite{HYO13}
\begin{equation}
 k = \frac{i}{r_e} -\frac{1}{r_e} \sqrt{\frac{2r_e}{a_s}-1}
=\frac{i}{r_e}\left(1-\sqrt{1-\frac{2r_e}{a_s}}\right),
\end{equation} 
where $a_s$ is a scattering length and $r_e$ is an effective range, respectively.
The effective range parameters deduced from the calculated phase shifts are
discussed in the following section. By assigning the values given in Table I
in the following section, the momentum
$k$ becomes $-0.079i$ for NLO13 and$-0.051i$ for NLO19. The
corresponding energy $E=\frac{\hbar^2}{2\mu_{\Lambda d}} k^2$, where $\mu_{\Lambda d}$
is a $\Lambda d$ reduced mass, is $-0.17$ MeV
for NLO13 and $-0.072$ MeV for NLO19.

It is noted that the imaginary part of the calculated phase shifts is small in both $J=1/2$
and $J=3/2$, which relates to the fact that the $^1 S_0$ channel
is not allowed in the final state.

\subsection{$\Lambda d$ effective range parameters}
\begin{figure}[thb]
\centering
\includegraphics[width=0.45\textwidth]{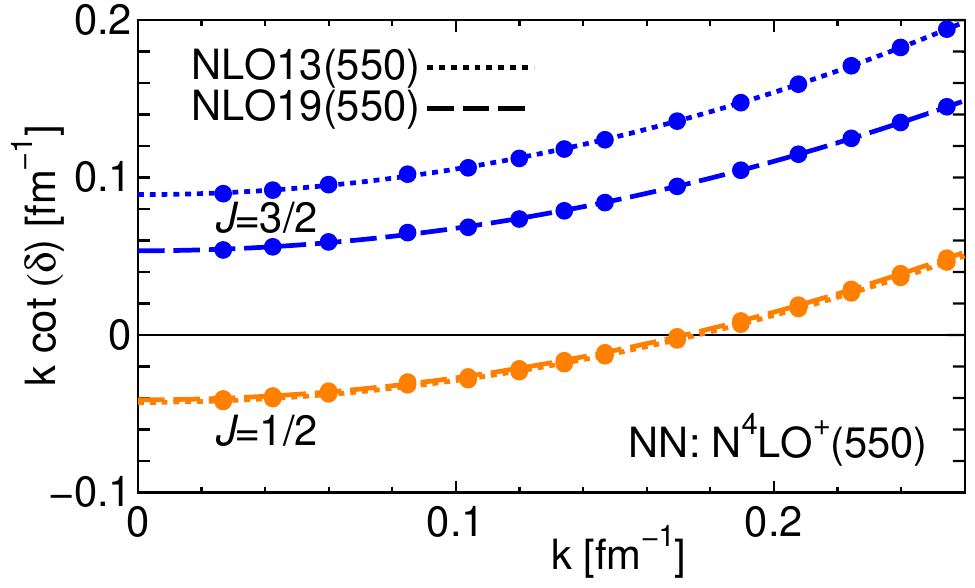}
\caption{$k\cot\delta$ as a function of the $\Lambda d$ relative momentum $k$.
The phase shift $\delta$ is real below the deuteron breakup momentum, $k \sim 0.28$ fm$^{-1}$.
The solid (dashed) curve is for the chiral NLO19 \cite{NLO19} (NLO13 \cite{NLO13}) $YN$ interaction
with the J{\"u}lich-Bonn group with the chiral N$^4$LO$^+$ $NN$ interaction \cite{RKE18}.
The solid and dashed curves are almost indistinguishable in the case of $J=1/2$.
}
\label{fig:eff}
\end{figure}

We estimate scattering length and effective range parameters by fitting
the calculated phase shifts, $k\cot \delta$ below the deuteron breakup
threshold, by a function $c_0+c_1 k^2+ c_2 k^4$.
Figure \ref{fig:eff} depicts the fit. Table \ref{tab:eff} tabulates the results of the scattering
length $a_s=-1/c_0$ and the effective range $r_e=2c_1$.
It is noted that these values are comparable to those of the preliminary results
from the measurement in heavy-ion collisions in Ref. \cite{HU24}:
$a_s^{J=1/2}=20_{-3}^{+3}$ fm, $r_e^{J=1/2}=3_{-1}^{+2}$ fm
and $a_s^{J=3/2}=-16_{-1}^{+2}$ fm, $r_e^{J=3/2}=2_{-1}^{+1}$ fm. 

The effective range parameters in the channel that bears a bound state are related to the
separation energy $B_\Lambda$ by a Bethe formula \cite{BETH49}, which gives
\begin{equation}
 B_\Lambda = \frac{\hbar^2}{\mu_{\Lambda d}r_e^2}\left\{ 1-\frac{r_e}{a_s}
 -\sqrt{1-\frac{2r_e}{a_s}} \right\}.
\end{equation}
Using the values in  Table \ref{tab:eff}, the expected $B_\Lambda$ becomes
$57.9$ keV for NLO13 and $53.3$ keV for NLO19. The Faddeev calculations \cite{KKM23}
for the bound hypertriton with the same $NN$ and $YN$ interactions predict
$B_\Lambda=79$ keV for NLO13 and $B_\Lambda=87$ keV for NLO19.
The magnitude of the difference is small but not negligible compared to the
small value of $B_\Lambda$. There are several sources of the difference.
The bound-state calculation in Ref. \cite{KKM23} explicitly includes
a $\Sigma NN$ component. However, the $\Sigma NN$ component is disregarded
in the present scattering calculation, though the
$\Lambda N$-$\Sigma N$ coupling is taken care of in calculating the $\Lambda N$ $t$ matrix. 
In addition, higher partial waves are included in the Faddeev three-body calculations.
It is noteworthy that the hypertriton separation energy obtained by Faddeev calculations
with ignoring the $\Sigma$ and higher partial waves in Faddeev components
is 73 keV for NLO13 and 76 keV for NLO19. These values
are closer to the estimates derived from the Bethe formula. The remaining discrepancy
may be attributed to the rearrangement effect of the deuteron wave function in forming the
hypertriton bound state.  

\begin{table}[t]
\begin{tabular}{|rc|cc|} \hline
  total spin& $YN$ int. & $a_s$ (fm) &$r_e$ (fm) \\ \hline
 $J=1/2$ & NLO13 & $\phantom{-}23.4$ & 2.77\\
              & NLO19 & $\phantom{-}24.3$ & 2.80\\ \hline
 $J=3/2$ & NLO13 & $-11.2$ & 3.25 \\
              & NLO19 & $-18.7$ & 2.87 \\ \hline
\end{tabular}
\caption{Scattering length and effective range parameters deduced
from the fit of the calculated phase shift presented in Fig. \ref{fig:eff}.
}
\label{tab:eff}
\end{table}

\subsection{$\Lambda d$ correlation function}
Explicit data from $\Lambda$-deuteron scattering experiments are not available.
An alternative way to probe the feature of the $\Lambda$d interaction has been developed
in heavy-ion collision experiments by measurement of $\Lambda$d correlations .
The $\Lambda$d momentum correlation function measured in experiments corresponds
to the following quantity \cite{FAB21,OMMH16}:
\begin{equation}
 C_{\Lambda d}^{J}(k)=1+4\pi \int_0^\infty r^2dr\; S_{12}(r) \{ |\psi_J(k;r)|^2-|j_0(kr)|^2\},
\end{equation}
where $j_0(kr)$ is a spherical Bessel function, and $\psi_J(k;r)$ is a $\Lambda$d scattering
wave function in the channel with the total spin of $J$. $S_{12}(r)$ is a source function that is
assumed to be a conventional Gaussian form with a range parameter $R$,
\begin{equation}
 S_{12}(r)=\frac{1}{(2\sqrt{\pi}R)^3} \exp (-\frac{1}{4R^2}r^2).
\end{equation}  
The $\Lambda$d wave function in coordinate space is constructed from half-off-shell
$T$ matrices obtained by solving the Faddeev equation.
The explicit expression is explained in Appendix C.
Realistic calculations of the correlation function taking into account the three-body
dynamics that the Faddeev wave functions yield is a future subject.

The total spin is unseparated in experiments. Therefore, the following spin-averaged
quantity is relevant when a comparison with the experimental data is made.
\begin{equation}
 C_{\Lambda d}(k)=\frac{1}{3} C_{\Lambda d}^{J=1/2}(k)+\frac{2}{3}C_{\Lambda d}^{J=3/2}(k).
\end{equation}  

\begin{figure}[t]
\centering
\includegraphics[width=0.4\textwidth]{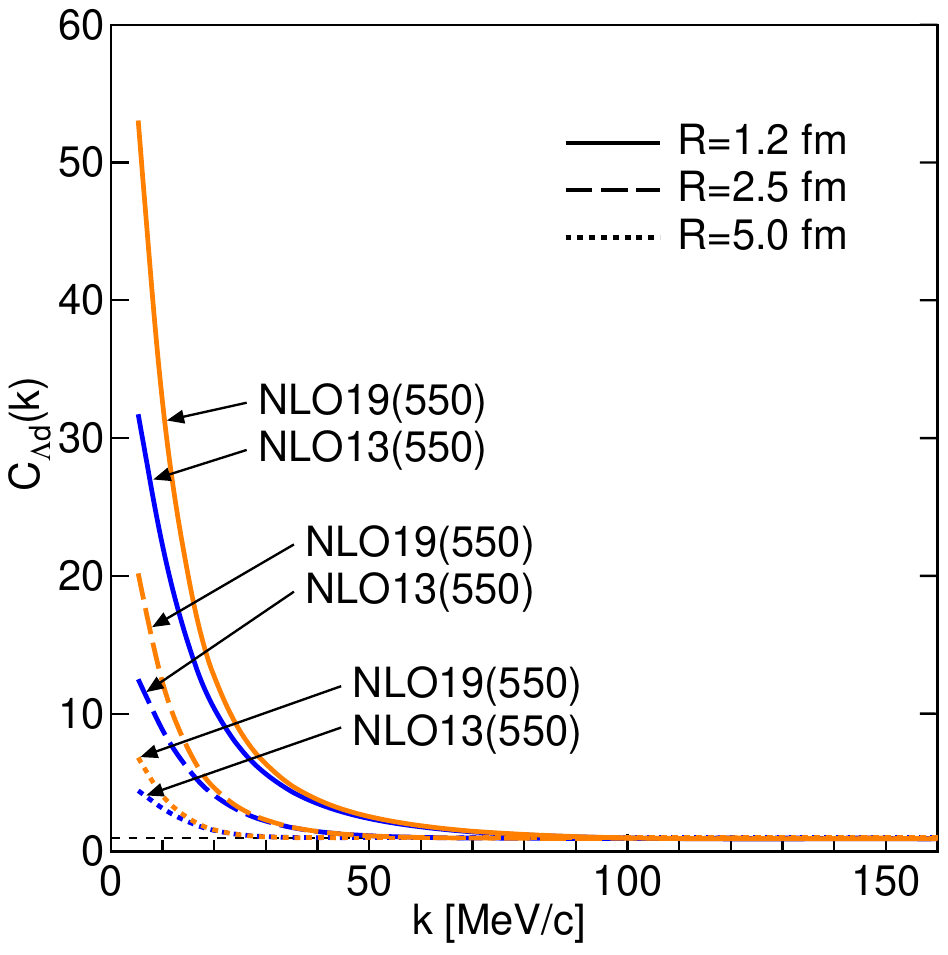}
\caption{Spin-averaged $\Lambda d$ correlation function as a function of the $\Lambda d$ relative
momentum $k$ on a vertical linear scale. The results using NLO13 \cite{NLO13}
and NLO19 $YN$ \cite{NLO19} interactions with N$^4$LO$^+$ $NN$ interactions \cite{RKE18}
are shown for three source ranges $R=1.2$, 2.5, and 5 fm. The cutoff scale is 550 MeV for both $NN$ and
$YN$ interactions.
\label{fig:avcr}
}
\end{figure}

Figure \ref{fig:avcr} displays the spin-averaged results of the NNLO13 and NNLO19 $YN$
interactions with the $NN$ interaction of N$^4$LO$^+$. 
The selection of the range parameter, $R=1.2$, 2.5, and 5 fm, follows that of Ref. \cite{HAID20}.
The difference between the results of NLO13 and NLO19 comes from $C_{\Lambda d}^{J=3/2}$.

\begin{figure}[H]
\centering
\includegraphics[width=0.4\textwidth]{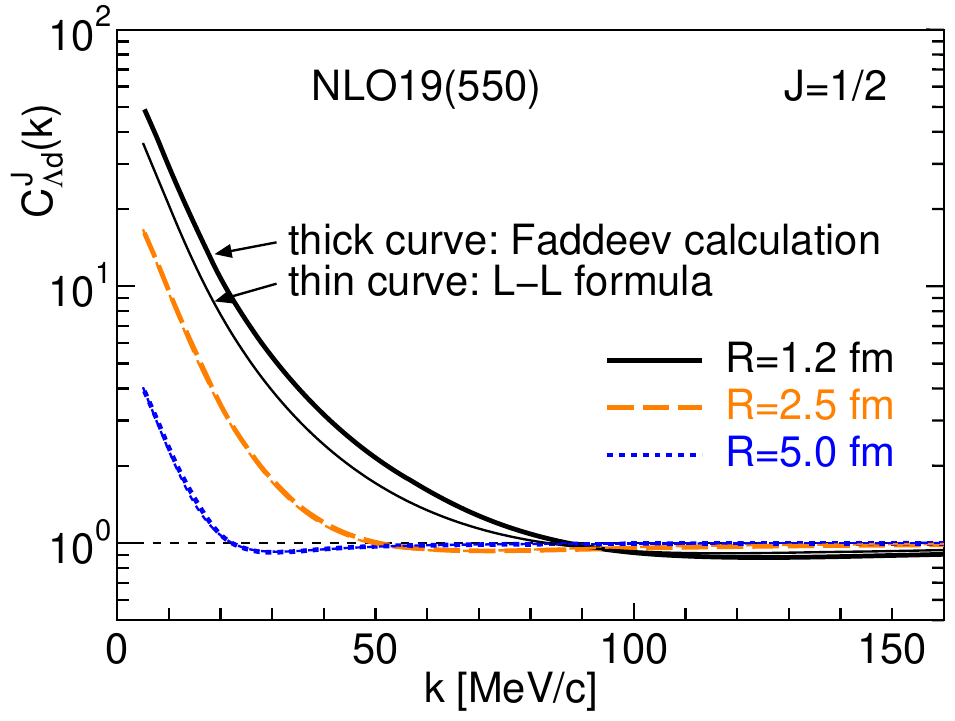}
\includegraphics[width=0.4\textwidth]{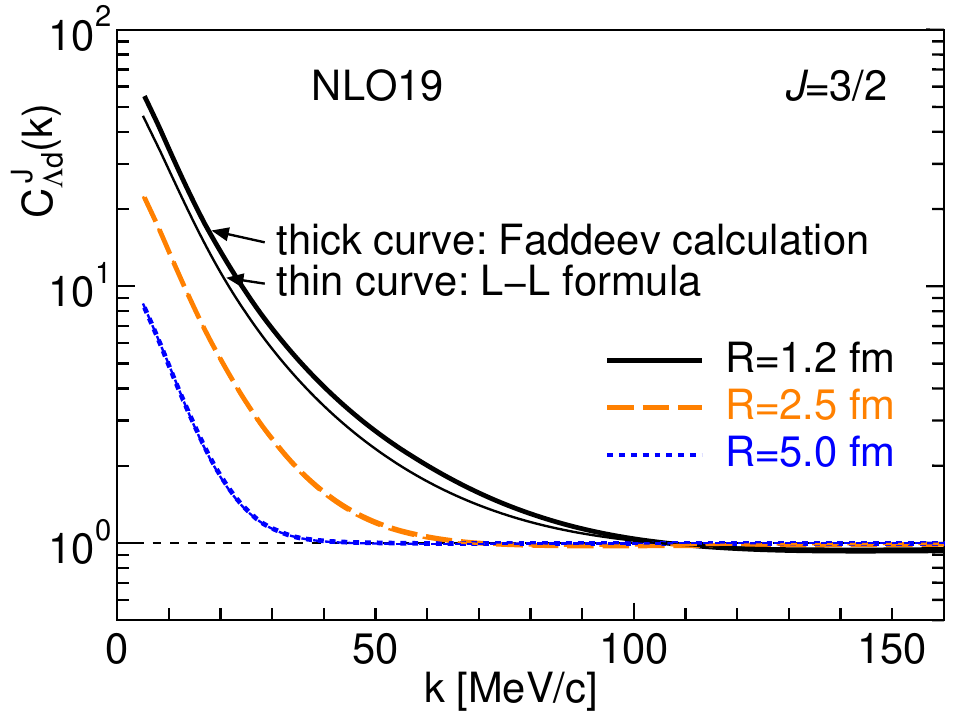}
\includegraphics[width=0.4\textwidth]{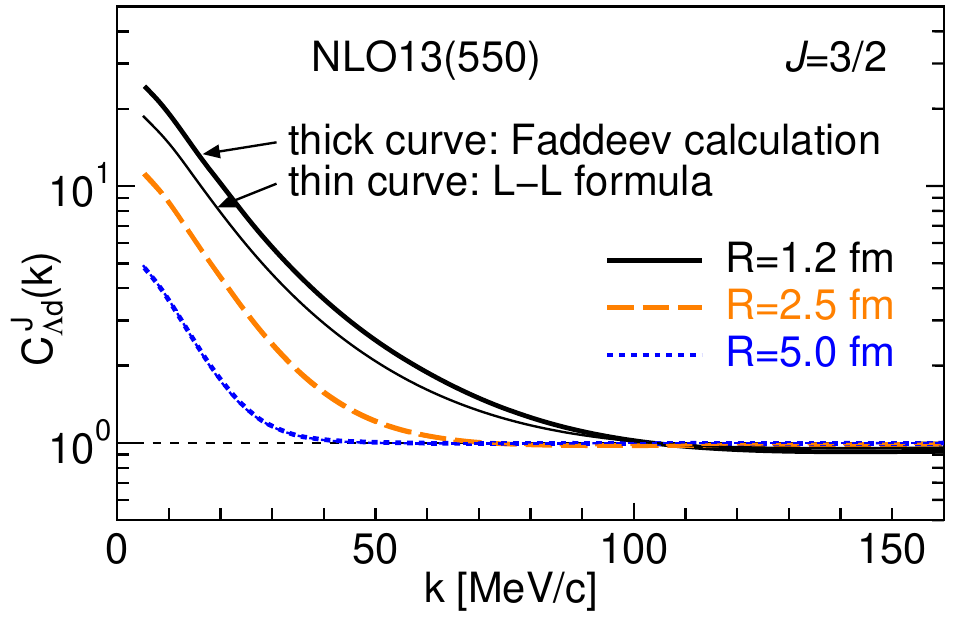}
\caption{$\Lambda d$ correlation function as a function of the $\Lambda d$ relative
momentum $k$ on a vertical logarithmic scale for the three choices of the source range $R$.
The thick curve is the result of the Faddeev calculation. The thin curve is the result
of the Lednick{\'y}-Lyuboshitz formula, given by Eq. (\ref{eq:app}), using the values in Table I.
The cutoff scale is 550 MeV for both $NN$ and $YN$ interactions.
The thick and thin curves overlap except for the case of $R=1.2$ fm in each panel.
The upper panel is the result of the NLO19 \cite{NLO19} $YN$ interaction for $J=1/2$,
in which the curves from NLO13 are not shown because
they almost overlap with those of NLO19. The middle and lower panels show
the results of NLO19 and NLO13 for $J=3/2$, respectively. $NN$ interactions are chiral
N$^4$LO$^+$ \cite{RKE18}.
\label{fig:corr}
}
\end{figure}

The individual momentum correlation function for $J=1/2$ and $J=3/2$ are shown in Fig. \ref{fig:corr}
on a vertical logarithmic scale.
The upper panel represents the correlation function for $J=1/2$, in which only the results of NLO19 are
depicted because NLO13 and NLO19 provide almost the same results as is expected from the
indistinguishability in the phase shifts,  The lower two panels show the result for $J=3/2$. In this case,
the correlation function of NLO19 is $2\sim 3$ times larger than that of NLO13.
Because the magnitude and weight of the $C_{\Lambda d}^{J=3/2}$ in which
only the $^3$S$_1$ $\Lambda$N interaction is relevant are larger than those of $C_{\Lambda d}^{J=1/2}$,
the experimental data can provide valuable information about the relative strength of the
$^1$S$_0$ and $^3$S$_1$ $\Lambda$N interactions.

The theoretical correlation function $C_{\Lambda d}^J$ has been approximated
by the following expression, which has been referred to as
the Lednick{\'y}-Lyuboshitz \cite{LL82} model formula:
\begin{align}
 C_{\Lambda d}^J \approx &\; 1+\frac{|f_J(k)|^2}{2R} F(r_e)
  +\frac{2\mbox{Re}f_J(k)}{\sqrt{\pi}R}F_1(x) \notag \\
 & -\frac{\mbox{Im}f_J(k)}{R}F_2(x),
\label{eq:app}
\end{align}
where $f_J$ is the scattering amplitude, $R$ is the range parameter,  and $x\equiv 2kR$. 
Three functions, i.e., $F$, $F_1$, and $F_2$, are given by
$F(r_e)=1-r_e/(2\sqrt{\pi}R)$, $F_1(x)= \int_0^x dt\:e^{t^2-x^2}/x$,
and $F_2(x)=(1-e^{-x^2})/x$. When $f_J$ is
approximated by the effective range parameters as
\begin{align}
 f_J=& \frac{e^{2i\delta_J}-1}{2ik} =\frac{1}{k\cot\delta_J -ik} \notag \\
 \approx & \frac{1}{-\frac{1}{a_s}+\frac{1}{2}r_e k^2-ik}.
\end{align} 
$C_{\Lambda d}^J$ is estimated by giving $R$, $a_s$, and $r_e$. It is instructive
to evaluate the approximated $C_{\Lambda d}^J$ using the effective range parameters
tabulated in Table \ref{tab:eff} and compare it with the correlation function obtained
by the $\Lambda$d relative wave function from the Faddeev calculation.

The thick curves in Fig. \ref{fig:corr} are the results of the Faddeev calculations
on a vertical logarithmic scale. The thin curve depicts
the correlation functions obtained by the effective range parameters.
Except for $R=1.2$ fm, the thick and thin curves are overlapping and
indistinguishable. The difference between the two curves in the case of $R=1.2$ fm is also
small. This result indicates that the approximation of the correlation function by Eq. (\ref{eq:app})
is dependable.

\section{Summary}
We describe $\Lambda$-deuteron elastic scattering in a Faddeev formulation for
low energies, up to 20 MeV, crossing the deuteron breakup threshold. Although direct
measurements of the scattering are not currently feasible, experimental information is
included in correlation functions that can be accessed through heavy-ion collision
measurements. The preliminary results were recently reported in Ref. \cite{HU24}.
The calculated phase shifts  are a valuable feature that elucidates the properties of the
underlying $YN$ interactions that are employed, including the effect of the relative strength
of the spin singlet and triplet parts in a few-body system.
It is also meaningful to investigate the implication of different parametrizations of the
$YN$ interactions in the scattering process of the $\Lambda$ hyperon on the deuteron. 

The numerical evaluation is more complex than that of nucleon-deuteron scattering
due to the mass difference between the $\Lambda$ hyperon and the nucleon.
For the sake of completeness,
an outline of the treatment of a deuteron pole and a moving singularity is provided
in Appendices A$-$C.
  
Based on the half-off-shell $t$ matrices obtained by solving the Faddeev
equation, $\Lambda d$ relative wave functions are constructed, and momentum
space $\Lambda d$ correlation functions are evaluated. The evaluated correlation
functions demonstrate the efficiency of the approximated
expression, given by Eq. (\ref{eq:app}),
although a slight deviation is observed when the source radius is small.
In Ref. \cite{MHS23}, the $\Lambda p$ correlation data are employed to constrain
parametrizing  $\Lambda N$ interactions. The $\Lambda d$ correlation data can provide additional
constraints through the use of Faddeev calculations.

Finally, it is noted that the present method can be straightforwardly applied to $\Xi d$ scattering,
which could help in studying the properties of $\Xi N$ interactions.

\bigskip
{\it Acknowledgments.}
We are grateful to K. Miyagawa for his valuable discussions and comments on this work.
This work is supported by JSPS KAKENHI Grants No. JP19K03849 and No. JP22K03597.

\begin{widetext}
\appendix
\section{Explicit equations in partial wave expansion}
Explicit equations in a partial wave expansion of Eqs. (6) and (7) are derived. We follow
the notation for the partial wave project state in Ref. \cite{GL96}.
\begin{align}
 & \langle \bp'|plm\rangle = \frac{\delta(p'-p)}{p'p} Y_{\ell m}(\hat{\bp'}), \hspace{2em}
 |p(ls)jm\rangle = \sum_\mu (\ell \mu s m-\mu |jm) |plm\rangle |s m-\mu\rangle, \\
 & \mbox{3-body state}\hspace{0.5em} |pq\alpha\rangle
 = |pq (\ell s)j (\lambda 1/2)I (jI)JM (t1/2)TM_T\rangle, \\
 & \langle p'q'\alpha'|pq\alpha\rangle =\frac{\delta(q'-q)}{q'q}\frac{\delta(p'-p)}{p'p} \delta_{\alpha'\alpha},
 \hspace{2em}\sum_\alpha\int_0^\infty p^2 dp \int_0^\infty q^2 dq \;|pq\alpha\rangle\langle pq\alpha| =I,
\end{align}
where $Y_{\ell m}$ stands for a spherical harmonics and $(\ell \mu s m-\mu |jm)$ represents
a Clebsch-Gordan coefficient. Inserting the identity operator, Eqs. (6) and (7) read
\begin{align}
  \langle p_3 q_3\alpha_3|T_3|\phi\rangle =& 2\sum_{\alpha_3'} \int_0^\infty {p_3'}^2 dp_3'
 \int_0^\infty {q_3'}^2 dq_3' \sum_{\alpha_2'} \int_0^\infty {p_2'}^2 dp_2' \int_0^\infty {q_2'}^2 dq_2' \notag \\
 & \times \langle p_3 q_3\alpha_3|t_3 G_0|p_3'q_3'\alpha_3'\rangle
\langle p_3'q_3'\alpha_3'|p_2'q_2'\alpha_2'\rangle \langle p_2'q_2'\alpha_2'|T_2|\phi\rangle, \\
\langle p_2 q_2\alpha_2|T_2|\phi\rangle =& \sum_{\alpha_2'} \int_0^\infty {p_2'}^2 dp_2'
 \int_0^\infty {q_2'}^2 dq_2' \sum_{\alpha_3'} \int_0^\infty {p_3'}^2 dp_3' \int_0^\infty {q_3'}^2 dq_3' \notag \\
 & \times \left\{ \langle p_2 q_2\alpha_2|t_2|p_2'q_2'\alpha_2'\rangle
 \langle p_2'q_2'\alpha_2'|p_3'q_3'\alpha_3'\rangle \langle p_3'q_3'\alpha_3'|\phi\rangle  \right. \notag\\
 &\left.+ \langle p_2 q_2\alpha_2|t_2 G_0|p_2'q_2'\alpha_2'\rangle
 \langle p_2'q_2'\alpha_2'|p_3'q_3'\alpha_3'\rangle \langle p_3'q_3'\alpha_3'|T_3|\phi\rangle \right\}\notag \\
 -& \sum_{\alpha_2'} \int_0^\infty {p_2'}^2 dp_2'
 \int_0^\infty {q_2'}^2 dq_2' \sum_{\alpha_2''} \int_0^\infty {p_2''}^2 dp_2'' \int_0^\infty {q_3''}^2 dq_2'' \notag\\
 & \times \langle p_2 q_2\alpha_2|t_2 |p_2'q_2'\alpha_2'\rangle
 \langle p_2'q_2'\alpha_2'|P_{12}G_0|p_2''q_2''\alpha_2''\rangle \langle p_2''q_2''\alpha_2''|T_2|\phi\rangle,
\end{align}
where the antisymmetric property of the state $\langle p_2'q_2'\alpha_2'|$ under the $P_{12}$ operation is used. 

The matrix elements of $t_3 G_0$, $t_2$, and $t_2 G_0$ are explicitly wriiten as
\begin{align}
 \langle p_3 q_3\alpha_3|t_3 G_0|p_3'q_3'\alpha_3'\rangle =&
 \frac{\delta(q_3-q_3')}{q_3q_3'} \langle p_3 \alpha_3|t_3(E-h_{\Lambda(NN)}(q_3)) |p_3'\alpha_3'\rangle
 \frac{1}{E-h_{NN}(p_3')-h_{\Lambda(NN)}(q_3)+i\epsilon},\\
 \langle p_2 q_2\alpha_2|t_2|p_2'q_2'\alpha_2'\rangle =&\frac{\delta(q_2-q_2')}{q_2q_2'}
 \langle p_2 \alpha_2|t_2(E-h_{N(\Lambda N)}(q_2)) |p_2'\alpha_2'\rangle, \\
 \langle p_2 q_2\alpha_3|t_2 G_0|p_2'q_2'\alpha_2'\rangle =&
 \frac{\delta(q_2-q_2')}{q_2q_2'} \langle p_2 \alpha_2|t_2(E-h_{N(\Lambda N)}(q_2)) |p_2'\alpha_2'\rangle
 \frac{1}{E-h_{\Lambda N}(p_2')-h_{N(\Lambda N)}(q_2)+i\epsilon},
\end{align}
where $h_{NN}(p_3')=\frac{\hbar^2}{2\mu_{NN}}{p_3'}^2$ with $\mu_{NN}=\frac{1}{2}m_N$,
$h_{\Lambda(NN)}(q_3)=\frac{\hbar^2}{2\mu_{\Lambda(NN)}}q_3^2$ with
$\mu_{\Lambda(NN)}=\frac{2m_Nm_\Lambda}{2m_N+m_\Lambda}$,
$h_{\Lambda N}(p_2')=\frac{\hbar^2}{2\mu_{\Lambda N}}{p_2'}^2$ with
$\mu_{N(\Lambda N)}=\frac{m_N m_\Lambda}{m_N+m_\Lambda}$,
and $h_{N(\Lambda N)}(q_2)=\frac{\hbar^2}{2\mu_{N(\Lambda N)}}q_2^2$ with
$\mu_{N(\Lambda N)}=\frac{m_N(m_N+m_\Lambda)}{2m_N+m_\Lambda}$. The matrix elements of
the permutation of Jacobi momenta have the following form:
\begin{align}
 \langle p_3'q_3'\alpha_3'|p_2'q_2'\alpha_2'\rangle=&\int_{-1}^1 d\cos \theta \;G_{\alpha_3' \alpha_2'}^{(2)}
 (q_3',q_2',\cos\theta)\frac{1}{{p_3'}^{\ell_{\alpha_3'}} {p_2'}^{\ell_{\alpha_2'}} }
 \frac{\delta(p_3'-\pi_3)\delta(p_2'-\pi_2)}{{p_3'}^2 {p_2'}^2}, \\
  \langle p_2'q_2'\alpha_2'|p_3'q_3'\alpha_3'\rangle=&\int_{-1}^1 d\cos \theta \;G_{\alpha_2' \alpha_3'}^{(1)}
 (q_2',q_3',\cos\theta)\frac{1}{{p_2'}^{\ell_{\alpha_2'}} {p_3'}^{\ell_{\alpha_3'}} }
 \frac{\delta(p_2'-\pi_2')\delta(p_3'-\pi_3')}{{p_2'}^2 {p_3'}^2},\\
 \langle p_2'q_2'\alpha_2'|P_{12}G_0|p_2''q_2''\alpha_2''\rangle=&
 \int_{-1}^1 d\cos \theta \;G_{\alpha_2' \alpha_2''}^{(3)} (q_2',q_2'',\cos\theta)
 \frac{1}{{p_2'}^{\ell_{\alpha_2'}} {p_2''}^{\ell_{\alpha_2''}} }
 \frac{\delta(p_2'-\pi_2'')\delta(p_2''-\pi_2''')}{{p_2'}^2 {p_2''}^2},
\end{align}
where $\cos\theta=\cos\widehat{\bq_3' \bq_2'}$ in Eq. (A9),
$\cos\theta=\cos\widehat{\bq_2' \bq_3'}$ in Eq. (A10), and
$\cos\theta=\cos\widehat{\bq_2' \bq_2''}$ in Eq. (A11), respectively.
Various momenta in the above equations are defined as follows:
\begin{align}
 \pi_3=& [q_2'^2 +r_{NN}^2 {q_3}^2 +2r_{NN} q_2' q_3'\cos\theta]^{1/2},
  & \pi_2=[{q_3}^2 +r_{\Lambda N}^2 q_2'^2 +2r_{\Lambda N} q_2' q_3'\cos\theta]^{1/2}, \notag \\
 \pi_{30}=& [q_2'^2 +r_{NN}^2 q_0^2 +2r_{NN} q_2' q_0\cos\theta]^{1/2},
  & \pi_{20}=[{q_0}^2 +r_{\Lambda N}^2 q_2'^2 +2r_{\Lambda N} q_2' q_0\cos\theta]^{1/2},\notag \\
 \pi_3'=& [q_2^2 +r_{NN}^2 {q_3'}^2 +2r_{NN} q_2 q_3'\cos\theta]^{1/2},
  & \pi_2'=[{q_3'}^2 +r_{\Lambda N}^2 q_2^2 +2r_{\Lambda N} q_2 q_3'\cos\theta]^{1/2},\notag \\
  \pi_2''=& [q_2''^2 +r_{N\Lambda}^2 q_2^2 +2r_{N\Lambda} q_2 q_2''\cos\theta]^{1/2},
  & \pi_2'''=[q_2^2 +r_{N\Lambda}^2 q_2''^2 +2r_{N\Lambda} q_2 q_2''\cos\theta]^{1/2},\notag \\
 \pi_{30}'=& [q_2^2 +r_{NN}^2 q_0^2 +2r_{NN} q_2 q_0\cos\theta]^{1/2},
 & \pi_{20}'= [{q_0}^2 +r_{\Lambda N}^2 q_2^2 +2r_{\Lambda N} q_2 q_0\cos\theta]^{1/2},\notag \\
  \pi_{3i}'=& [{q_2}^2 +r_{NN}^2 q_i^2 +2r_{NN} q_2 q_i\cos\theta]^{1/2},
 & \pi_{2i}'= [{q_i}^2 +r_{\Lambda N}^2 q_2^2 +2r_{\Lambda N} q_2 q_i\cos\theta]^{1/2},
\label{eq:mom}
\end{align}
where $r_{NN}=\frac{1}{2}$, $r_{N\Lambda}=\frac{m_N}{m_N+m_\Lambda}$,
and $r_{\Lambda N}=\frac{m_\Lambda}{m_N+m_\Lambda}$, respectively. 
Before transforming Eqs. (A4) and (A5) further, it is necessary to explain the treatment of
the deuteron pole in the $NN$ $t$ matrix. 

\section{Treatment of deuteron pole}
In the Faddeev equation, the $NN$ $t$ matrix depends on the
momentum $q_3$ of the third particle $\Lambda$.
\begin{equation}
 t(q_3) = v + v\frac{1}{E-h_{\Lambda(NN)}(q_3) -h_{NN}(p_3) +i\epsilon} t(q_3)
          = v + v\frac{1}{E-h_{\Lambda(NN)}(q_3) -h_{NN}(p_3) -v +i\epsilon} v,
\label{eq:tm}
\end{equation}
where $E$ is the total energy, and $v$ is a two-body $NN$ interaction.
It is helpful to introduce a spectral decomposition to investigate the structure of the $t$ matrix.
Assuming that there is one bound state $|d\rangle$ with its energy $-|e_d|$ for the Hamiltonian
$H=\frac{\hbar^2}{2\mu_{NN}}p_3^2 +v$ and denoting the eigenstate in the continuum
with its momentum $\bk$ by $|\psi(\bk)\rangle$, the completeness relation reads
\begin{align}
 \int \frac{d\bk}{(2\pi)^3} |\psi(\bk)\rangle \langle \psi(\bk)| + |d\rangle \langle d|.
\end{align} 
By inserting this relation into Eq. (\ref{eq:tm}), the following expression is obtained.
\begin{align}
 t(q_3) =& v + \int \frac{d\bk}{(2\pi)^3} \frac{v|\psi(\bk)\rangle \langle \psi(\bk)|v}
 {E-h_{\Lambda(NN)}(q_3) -h_{NN}(k) -v +i\epsilon}
 +\frac{v|d\rangle \langle d|v}{E-h_{\Lambda(NN)}(q_3) +|e_d| +i\epsilon}.
\label{eq:spt}
\end{align}
The $t$ matrix is solved numerically. The expression of Eq. (\ref{eq:spt}) is used for
the prescription to treat the singularity at the deuteron pole position. The denominator of the
second term has a pole for the momentum $k$ when $E-h_{\Lambda(NN)}(q_3) >0$.
This feature is known as a moving pole. The third term of Eq. (\ref{eq:spt}) has a pole that
appears in the $t$ matrix of the deuteron channel when $E_{cm}\equiv E+|e_d| >0$.
The singularity is written as
\begin{equation}
 \frac{v|d\rangle \langle d|v}{E-h_{\Lambda(NN)}(q_3) +|e_d| +i\epsilon}=
 \mbox{P}\frac{v|d\rangle \langle d|v}{E-h_{\Lambda(NN)}(q_3) +|e_d|
 +i\epsilon}-i\pi \delta(E-h_{\Lambda(NN)}(q_3) +|e_d|)v|d\rangle \langle d|v.
\label{eq:sg}
\end{equation}
Because the $t$-matrix solved numerically does not contain the $\delta$-function part,
it has to be added separately as in Ref. \cite{FF10}. The $\delta$-function part can be written
as follows:
\begin{equation}
 -i\pi \delta(E-h_{\Lambda(NN)}(q_3) +|e_d|)v|\;d\rangle \langle d|v
 = -i \frac{\pi}{2q_0}\frac{2\mu_{\Lambda (NN)}}{\hbar^2}\delta(q_3-q_0) v|d\rangle \langle d|v,
\end{equation}
where $q_0$ defined by $E+|e_d|\equiv \frac{\hbar^2}{2\mu_{\Lambda(NN)}}q_0^2$ is
introduced.

It is convenient to treat the $\delta$-function part separately in the set of Faddeev equations. 
Because the bound-state
wave function is known, and $\langle p_3|v|d\rangle =\left\{ e_d-2\frac{\hbar^2}{\mu_{NN}}p_3^2\right\}
\langle p_3|d\rangle$, the matrix element of the numerator of the third term is rewritten as follows:
\begin{align}
 \langle p_3|v|d\rangle \langle d|v| p_3'\rangle=\left\{ e_d-\frac{\hbar^2}{2\mu_{NN}}p_3^2\right\}
 \left\{ e_d-\frac{\hbar^2}{2\mu_{NN}}{p_3'}^2\right\}\langle p_3|d\rangle \langle d|p_3'\rangle,
\end{align}
where $\langle p_3|d\rangle$ is a bound-state (deuteron) wave function in momentum space.

It is convenient to separate the contribution
of the $\delta$-function part of Eq. (\ref{eq:sg}) and express $T_3$ as $T_3=T_3^R+iT_3^I$.
It is noted that both $T_3^R$ and $T_3^I$ are complex.
Then, Eqs. (A4) and (A5) are expressed as follows. 
\begin{align}
 \langle p_3 q_3\alpha_3|T_3^R|\phi\rangle
 =& 2\sum_{\alpha_3'}\sum_{\alpha_2'} \int q_2'^2 dq_2' \int_{-1}^1 d\cos\theta
 \langle p_3 \alpha_{3'}|t_3(E-h_{\Lambda(NN)}(q_3)) |\pi_3\alpha_3' \rangle
 \frac{1}{E-h_{NN}(\pi_3) -h_{\Lambda(NN)}(q_3)+i\epsilon} \notag \\
 & \times  G_{\alpha_3'\alpha_2'}^{(2)} (q_3,q_2',\cos\theta)
 \frac{1}{{\pi_3}^{\ell_{\alpha_3'}} {\pi_2}^{\ell_{\alpha_2'}}}
 \langle \pi_2 q_2'\alpha_2'|T_2|\phi\rangle, \\
  \langle p_3 q_0\alpha_3|T_3^I|\phi\rangle
= &  -\delta_{\alpha_3 \alpha_d} \left(\epsilon_d-h_{NN}(p_3) \right)
 \frac{2\mu_{\Lambda(NN)}}{\hbar^2}\pi q_0
 \sum_{\alpha_d'}\sum_{\alpha_2'} \int q_2'^2 dq_2' \int_{-1}^1 d\cos\theta \;
  \langle p_3\alpha_d|\psi_d\rangle \langle \psi_d|\pi_{30} \alpha_d'\rangle \notag \\
 &\times G_{\alpha_3'\alpha_2'}^{(2)} (q_0,q_2',\cos\theta)
 \frac{1}{{\pi_{30}}^{\ell_{\alpha_3'}} {\pi_{20}}^{\ell_{\alpha_2'}}}
 \langle \pi_{20} {q_2}'\alpha_2'|T_2|\phi\rangle, \\
 \langle p_2 q_2\alpha_2|T_2|\phi\rangle =&\sum_{\alpha_2'}\sum_{\alpha_d'}
  \int_{-1}^1 d\cos\theta \langle p_2 \alpha_2|t_2 |\pi_{2i}'\alpha_{2i}'\rangle \;
  G_{\alpha_2'\alpha_d'}^{(1)} (q_2,q_i,\cos\theta)
 \frac{1}{{\pi_{2i}'}^{\ell_{\alpha_2'}} {\pi_{3i}'}^{\ell_{\alpha_d'}}}
  \langle \pi_{3i}'\alpha_d'|\psi_d\rangle \notag \\
 + & \sum_{\alpha_2'}\sum_{\alpha_3'} \int q_3'^2 dq_3' \int_{-1}^1 d\cos\theta\;
 \langle p_2 \alpha_2|t_2 |\pi_2'\alpha_2'\rangle\;
 \frac{1}{E-h_{\Lambda N}(\pi_2')
 -\frac{\hbar^2}{2\mu_{N(\Lambda N)}}q_2^2+i\epsilon} \notag \\
 & \times G_{\alpha_2'\alpha_3'}^{(1)} (q_2,q_3',\cos\theta)
  \frac{1}{{\pi_2'}^{\ell_{\alpha_2'}} {\pi_3'}^{\ell_{\alpha_3'}}}
 \langle \pi_3'q_3'\alpha_3'|T_3^R|\phi\rangle \notag \\
 + &  i \sum_{\alpha_2'}\sum_{\alpha_d} \int_{-1}^1 d\cos\theta\;
 \langle p_2 \alpha_2|t_2 |\pi_{20}'\alpha_2'\rangle
  G_{\alpha_2'\alpha_3'}^{(1)} (q_2,q_0,\cos\theta)
  \frac{1}{{\pi_{20}'}^{\ell_{\alpha_2'}} {\pi_{30}'}^{\ell_{\alpha_3'}}}
  \langle \pi_{30}'q_0\alpha_3'|\tilde{T}_3^I|\phi\rangle \notag \\
 - & \sum_{\alpha_2'} \sum_{\alpha_2''} \int q_2''^2 dq_2'' \int_{-1}^1 d\cos\theta\;
 \langle p_2 \alpha_2|t_2 |\pi_2''\alpha_2'\rangle
 \; G_{\alpha_2'\alpha_2''}^{(3)} (q_2,q_2'',\cos\theta)
  \frac{1}{{\pi_2''}^{\ell_{\alpha_2'}} {\pi_2'''}^{\ell_{\alpha_2''}}} \notag \\
 & \times \frac{1}{E-h_{\Lambda N}(\pi_2''')
 -h_{N(\Lambda N)}(q_2'')+i\epsilon}
 \langle \pi_2'''q_2''\alpha_2''|T_2|\phi\rangle,
\end{align}
In Eq. (B7),  $\langle p_3 \alpha_{3'}|t_3(E-h_{\Lambda(NN)}(q_3)) |\pi_3\alpha_3' \rangle$
in the deuteron channel is understood as
\begin{equation}
 \langle p_3|t_3(E-h_{\Lambda(NN)}(q_3)) |\pi_3\rangle-\frac{2\mu_{\Lambda(NN)}}{\hbar^2}
\frac{\langle p_3|v|d\rangle \langle d|v|\pi_3\rangle}{q_0^2-q_3^2}
 +\mbox{P}\frac{2\mu_{\Lambda(NN)}}{\hbar^2}
\frac{\langle p_3|v|d\rangle \langle d|v|\pi_3\rangle}{q_0^2-q_3^2}.
\end{equation}
The divergent behavior of the $t_3$-matrix element around the pole position is removed
by the second term.
When $T_3^R$ is inserted in Eq. (B9), the principal value of $q_3'$ integration is treated
by the standard subtraction method.

When $E=E_{cm}+e_d >0$, there is a notorious problem of moving singularities \cite{GL96}
in which logarithmic singularities appear. We use numerical techniques in the literature
\cite{FF10,LIU05}, but the actual calculations become intricate because of the mass
difference between the $\Lambda$ hyperon and the nucleon. 

\section{Crescent area and logarithmic singularity}
Three different types of the crescent area in which the logarithmic singularity associated with
the so-called moving pole of the Green function exists
in the case of $E=E_{cm}+e_d >0$ appear, which are depicted in Fig. \ref{fig:cre}. Various reference
points shown in these figures are defined as follows:
\begin{align}
 q_{2,m}'=&q_{2,m}=\sqrt{\frac{2\mu_{N(\Lambda N)}}{\hbar^2}E},\hspace{1em}
  q_{2,z}'=q_{2,az}=\sqrt{\frac{2\mu_{NN}}{\hbar^2}E},\hspace{1em}
  q_{2,t}'=q_{2,at}=\sqrt{\frac{\mu_{N(\Lambda N)}}{\hbar^2}Er_{\Lambda N}}
   =r_{NN}\sqrt{\frac{2\mu_{\Lambda(NN)}}{\hbar^2}E}, \notag \\
q_{3,m}=&q_{3,am}'=\sqrt{\frac{2\mu_{\Lambda(NN)}}{\hbar^2}E},\hspace{1em}
  q_{3,z}=q_{3,az}'=q_{2,bz}'=q_{2,bz}=\sqrt{\frac{2\mu_{\Lambda N}}{\hbar^2}E},\hspace{1em}
  q_{3,t}=\sqrt{\frac{\mu_{\Lambda(NN)}}{\hbar^2}Er_{\Lambda N}}, \notag \\
  q_{3,at}'=&\sqrt{\frac{\mu_{\Lambda(NN)}}{\hbar^2}Er_{\Lambda N}}
 =r_{\Lambda N}\sqrt{\frac{2\mu_{N(\Lambda N)}}{\hbar^2}E},\hspace{1em}
   q_{2,bt}'=q_{2,bt}=r_{N\Lambda} \sqrt{\frac{2\mu_{N(\Lambda N)}}{\hbar^2}E}.
\end{align}
The mesh points for the $q_3$ momentum are set by separating them into the
following three intervals:
\begin{equation}
  [0,q_{3,t}], [q_{3,t},q_{3,z}], [q_{3,z},q_{3,m}].
\end{equation}
Because of $q_{2,at}\neq q_{2,bt}$ and $q_{2,az}\neq q_{2,bz}$, the mesh points for
the $q_2$ momentum are set by separating them into the following five intervals:
\begin{equation}
  [0,q_{2,bt}], [q_{2,bt},q_{2,at}], [q_{2,at},q_{2,az}], [q_{2,az},q_{2,bz}], [q_{2,bz},q_{2,m}].
\end{equation}
The mesh points near $q_{2,m}$, $q_{2,az}$, and $q_{2,bz}$ are prepared by changing
the variable to $\sqrt{q_{2,m}^2-q_2^2}$ \textit{etc.} The Gauss-Legendre quadrature
is used in each interval. The sufficiently fine mesh points are prepared around the deuteron
pole $q_0$ that is larger $q_{3,m}$. In solving Eqs. (A7)$-$(A9), the momenta given
in (\ref{eq:mom}), in general, do not coincide with the prepared mesh points. This problem
is circumvented by the technique of using a cubic Spline interpolation. The notorious
logarithmic singularities in the crescent area are treated by the method described
in detail in Refs. \cite{FF10} and \cite{LIU05}.

\begin{figure}[t]
\centering
\includegraphics[width=0.32\textwidth]{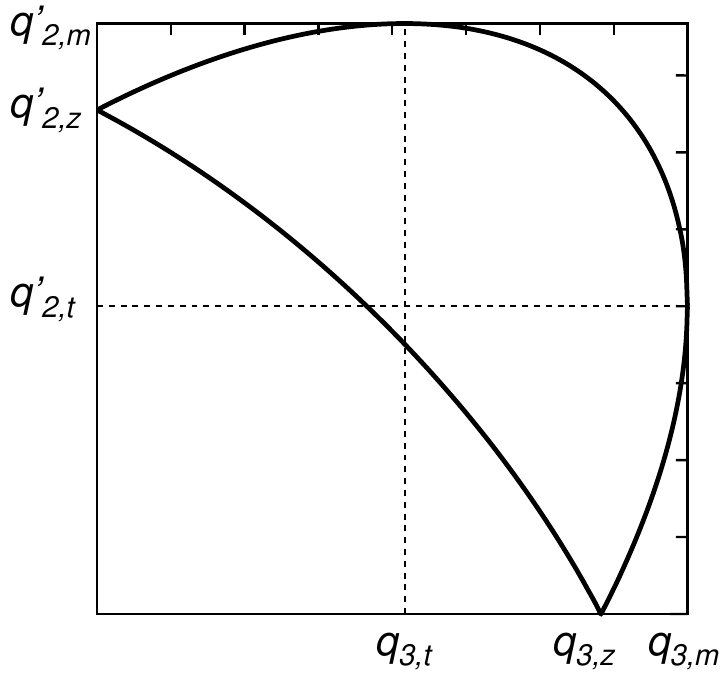}
\includegraphics[width=0.33\textwidth]{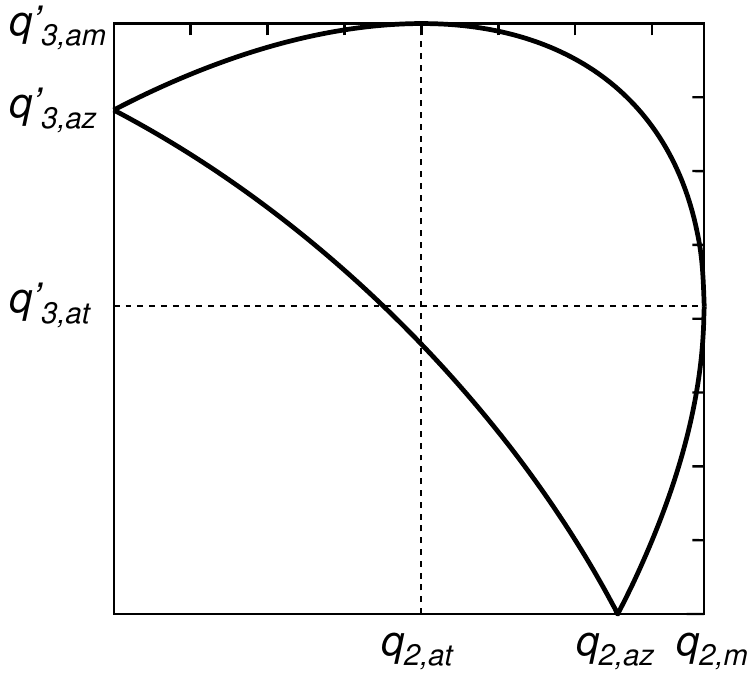}
\includegraphics[width=0.33\textwidth]{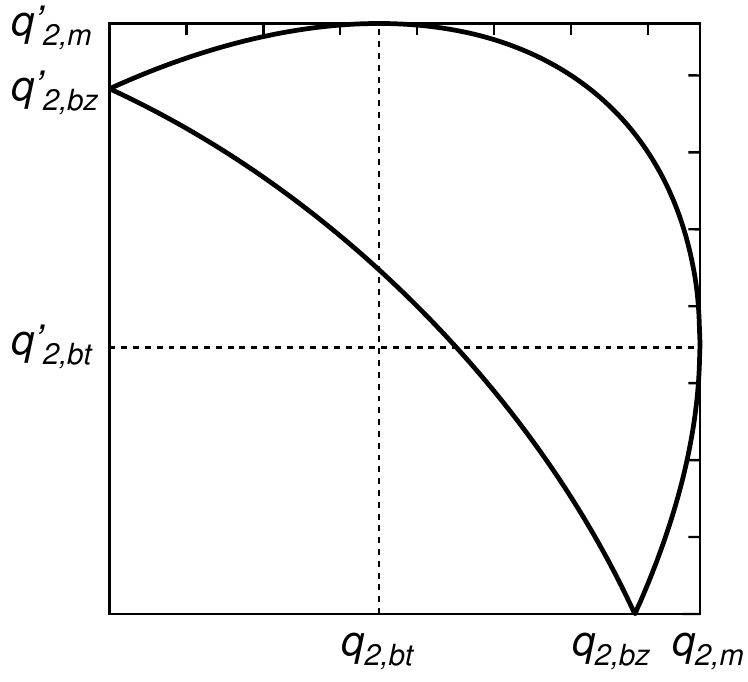}
\caption{Three types of the so-called crescent area in which a careful treatment of
the logarithmic singularities is necessary in the case of $E=E_{cm}+e_d >0$.
}
\label{fig:cre}
\end{figure}

\section{$\Lambda$-deuteron wave function in $r$-space}
Half-off-shell $T_2$ matrices in Eq. (\ref{eq:pe2}) determine the $\Lambda d$ relative
wave function in $r$ space. A partial wave component is obtained by
\begin{equation}
 \psi_{k,\ell}^+ (r)=j_{\ell}(kr) +\frac{2\mu_{\Lambda d}}{\hbar^2}
 \int_0^\infty k'^2dk'\; \frac{j_\ell (k'r) T_{2,\ell} (k',k)}{k^2+i\epsilon -k'^2},
\end{equation}
where the tensor coupling is suppressed for simplicity.
The integration is evaluated numerically by using a subtraction prescription,
\begin{equation}
  \int_0^\infty k'^2dk'\; \frac{j_\ell (k'r) T_{2,\ell (k',k)}}{k^2+i\epsilon -k'^2}
= \int_0^\infty dk' \frac{k'^2 j_\ell(k'r)T_{2,\ell(k',k)}-k^2 j_\ell(kr)T_{2,\ell(k,k)}}{k^2-k'^2}
 -i\frac{\pi k}{2} j_\ell(kr)T_{2,\ell}(k,k).
\end{equation}
 
\end{widetext}

\end{document}